\documentclass[twocolumn,showpacs,preprintnumbers,superscriptaddress,pre]{revtex4}

\usepackage{graphicx}
\usepackage{dcolumn}
\usepackage{bm}
\usepackage{color}
\usepackage{amssymb,amsmath,amsfonts}

\usepackage{CJK}

\begin{document}
\begin{CJK*}{GBK}{song}

\title{Modeling correlated human dynamics}

\author{Peng Wang}
\affiliation{Department of Modern Physics, University of Science
and Technology of China, Hefei 230026, China}
\affiliation{Department of Physics, University of Fribourg,
Fribourg 1700, Switzerland}
\author{Tao Zhou}
\affiliation{Department of Modern Physics, University of Science
and Technology of China, Hefei 230026, China} \affiliation{Web
Sciences Center, University of Electronic Science and Technology
of China, Chengdu 610054, China}
\author{Xiao-Pu Han}
\affiliation{Department of Modern Physics, University of Science and
Technology of China, Hefei 230026, China}
\author{Bing-Hong Wang}
\affiliation{Department of Modern Physics, University of Science and
Technology of China, Hefei 230026, China} \affiliation{The Research
Center for Complex System Science, University of Shanghai for
Science and Technology and Shanghai Academy of System Science,
Shanghai, 200093 China}
\date{\today}

\begin{abstract}
We empirically study the activity patterns of individual
blog-posting and find significant memory effects. The memory
coefficient [K.-I. Goh and A.-L. Barab\'asi, EPL {\bf 81}, 48002
(2008)] first decays in a power law and then turns to an
exponential form. Moreover, the inter-event time distribution
displays a heavy-tailed nature with power-law exponent dependent
on the activity. Our findings challenge the priority-queue model
[A.-L. Barab\'asi, Nature {\bf 435}, 207 (2005)] that can not
reproduce the memory effects or the activity-dependent
distributions. We think there is another kind of human activity
patterns driven by personal interests and characterized by strong
memory effects. Accordingly, we propose a simple model based on
temporal preference, which can well reproduce both the
heavy-tailed nature and the strong memory effects. This work helps
in understanding both the temporal regularities and the
predictability of human behaviors.
\end{abstract}

\pacs{89.75.Da, 02.50.-r}

\maketitle

\end{CJK*}

\section{Introduction}
\label{S1:Intr}

Human actions underly many social, technological and economic
phenomena, and thus the quantitative understanding of human behavior
is very significant \cite{impact1,impact2}. Thanks to the
development of the information techniques, more and more electronic
records available from Internet may provide us insights into the
patterns of human behavior \cite{review1,review2}. In recent years,
examples of empirically studied human activities include
communication patterns of electronic mails \cite{dm7,dm1,dm2} and
surface mails \cite{dm2,dlett1,dlett2}, web surfing
\cite{dweb1,dweb2}, short message \cite{dmessage1}, movie ratings
\cite{activity1}, online game \cite{dgame1,dgame2}. The main result,
arising from all these studies, concerns the heavy-tailed natures of
human activity: the inter-event and/or response times follow a
power-law-like distribution at the level of both population and
individual.

A possible explanation of the heavy-tailed nature is the
priority-queue model firstly introduced by Barab\'asi
\cite{dm7,dm8}, in which human behavior is primarily driven by
rational decision making. Another possible origin is the cascading
nonhomogeneous poisson process which emphasizes the external
factors such as circadian and weekly cycles \cite{m1,dm1,dm2}.
Although both models can give rise to heavy-tailed distribution of
inter-event and/or response times, the internal correlation of the
activities of human, as the most complex creature on earth, is
absent. However, in the common sense, our activities should
display memory effects since the personal tastes and interests are
known to have both the long-term consistence and short-term
burstiness. Long-term temporal memory in some human-initiated
systems has already been observed
\cite{correlation2,correlation3}. Moreover, the significant
potential predictability found in human mobility can be considered
as a complementary evidence of spatial memory of human activities
\cite{predict}. Actually, our daily activities can be roughly
divided into two classes: things we have to do and things we want
to do. Sending emails, making calls, submitting programmes in
Linux servers, printing papers can be seen as the first class,
which are important yet may not be interested to us. In contrast,
entertainment activities, such as listening to music, watching
movies and reading books, are driven by personal interests and
thus belong to the second class. Models considering adaptive
interests can to some extent reproduce the memory effects
\cite{m4,m2,m9}. Besides the memory effects, by extensive
empirical analyses on more than 10 real systems
\cite{dmessage1,activity1,activity2,activity3}, it is shown that
the individual activity (i.e., the frequency of actions of an
individual) plays an important role in determining the
distribution of inter-event time distribution: the larger the
activity the narrower the distribution. Both the memory effects
and the activity-dependent distributions can not be reproduced by
the priority-queue model with two universality classes \cite{dm8}.

In this paper, we empirically study the activity patterns of
individual blog-posting and find significant memory effects.
Moreover, the inter-event time distribution displays a
heavy-tailed nature with power-law exponent dependent on the
activity. We propose a simple model based on temporal preference,
which can well reproduce both the heavy-tailed nature and the
strong memory effects. This paper is organized as follows. In the
next section, we introduce the empirical observations, followed by
the model and simulation results in Section 3. We conclude this
work in Section 4 with some discussion about the relevance of our
work to the real human behavior.

\begin{figure}[htb]
\includegraphics[width=0.45\textwidth]{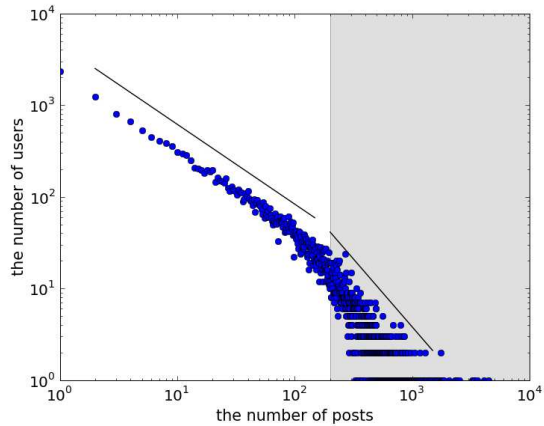}
\caption{\label{Fig:RI:PDF}(Color online)The distribution of the
number of posts. There are two scaling regimes, the exponents of
which are -1.48, -0.87. The part in the shadow correspond to the
users whose number of posts is greater than 200.}
\end{figure}

\begin{figure}[htb]
\includegraphics[width=0.4\textwidth]{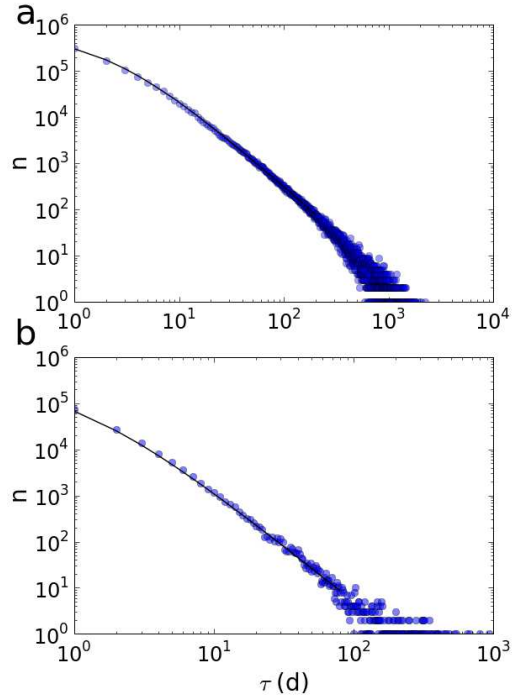}
\caption{\label{Fig:RI:PDF}(Color online) (a) The distribution of
interevent time of the whole population. $n$ is the number of
interevent times ($\tau$). We fit this distribution with the so
called "shifted power-law": $y\sim(x+a)^{-\beta}$\cite{book}. For
(a), the exponent $\beta=1.98$ and $a=2.1$. (b)The global interevent
time distribution of the uses whose number of posts is more than
600. the exponent $\beta=2.42$ and $a=1.0$. The average of
$Activity$ is 0.76 that would be why its exponent is larger than the
one of fig a.}
\end{figure}

\begin{figure}[htb]
\includegraphics[width=0.45\textwidth]{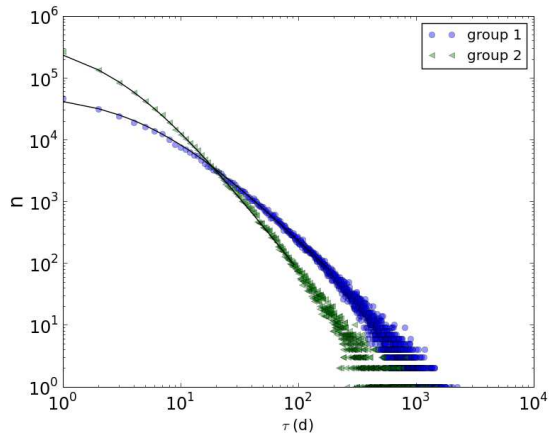}
\caption{\label{Fig:RI:PDF}(Color online)The distribution of
interevent time of the two groups. The fitting result is as follow:
for group 1, $\beta=1.86$ and $a=5.5$; for group 2, $\beta=2.47$ and
$a=3.0$. The exponent $\beta$ increase from 1.86 to 2.47 as the
$Activity$ from 0.04 to 0.64.}
\end{figure}

\section{Empirical Analysis}
Blog is a kind of so-called web2.0 application emerging in recent
years, in which people post some words, read and comment it each
other. For most ordinary bloggers(the user of blog), they post in
their own interest and treat it only as an amusement or an optional
way of communication with friends. This make their blogging unstable
and the frequency of it could be relatively low(often once one day).
Our data was collected from a campus blog website of Nanjing
university(http://bbs.nju.edu.cn/blogall). Most users are current or
former students and teachers of Nanjing university. As of
01/09/2009, there are 1627697 posts belonged to 20379 users in this
website. The first post is at 25/03/2003 when this blog established.
In fig 1, the distribution of the number of the post decays as
so-called double power law. The same result was also reported by
Grabowski\cite{blog1} who though that there are two groups of people
which clause two scaling regimes.

The heavy-tailed nature of the global distribution of interevent
times of all users is shown on the fig 2a. the exponent of this
distribution is -1.98 which is very close to the one in movie
rating\cite{activity1}and web activities on AOL and
Ebay\cite{activity2}. Figure 2b is the global interevent times
distribution of the users whose number of posts is more than 600.
One peculiarity of this distribution is that it has overmany long
intervals as we can see in the part of interevent time $\tau > 200$.
The similar feature also can be found in the distribution of single
user (seeing the insert of fig 4). But it is absent in fig 2a that
shows only mature users whose number of posts is larger have this
feature. We also can see the exponent of it is more than the one in
fig 2a. That actually is conformance to the dependence between the
exponent and $Activity$ which we will study below, since the users
who have more number of posts can often be more active.

\begin{figure}[htb]
\includegraphics[width=0.45\textwidth]{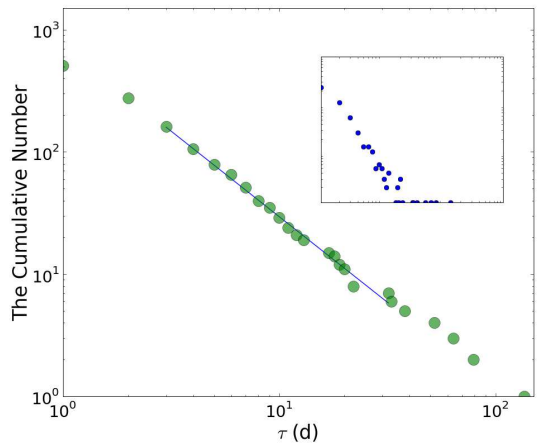}
\caption{\label{Fig:RI:PDF}(Color online) The cumulative
distribution of interevent times of one user. The distribution of
interevent times was shown in the inset of it. the exponents of the
cumulative one of this user is 1.40. Correspondingly, the one of
distribution of this user are 2.40. }
\end{figure}

\begin{figure*}[ht]
\begin{center}
\centerline{\includegraphics[width=.8\textwidth]{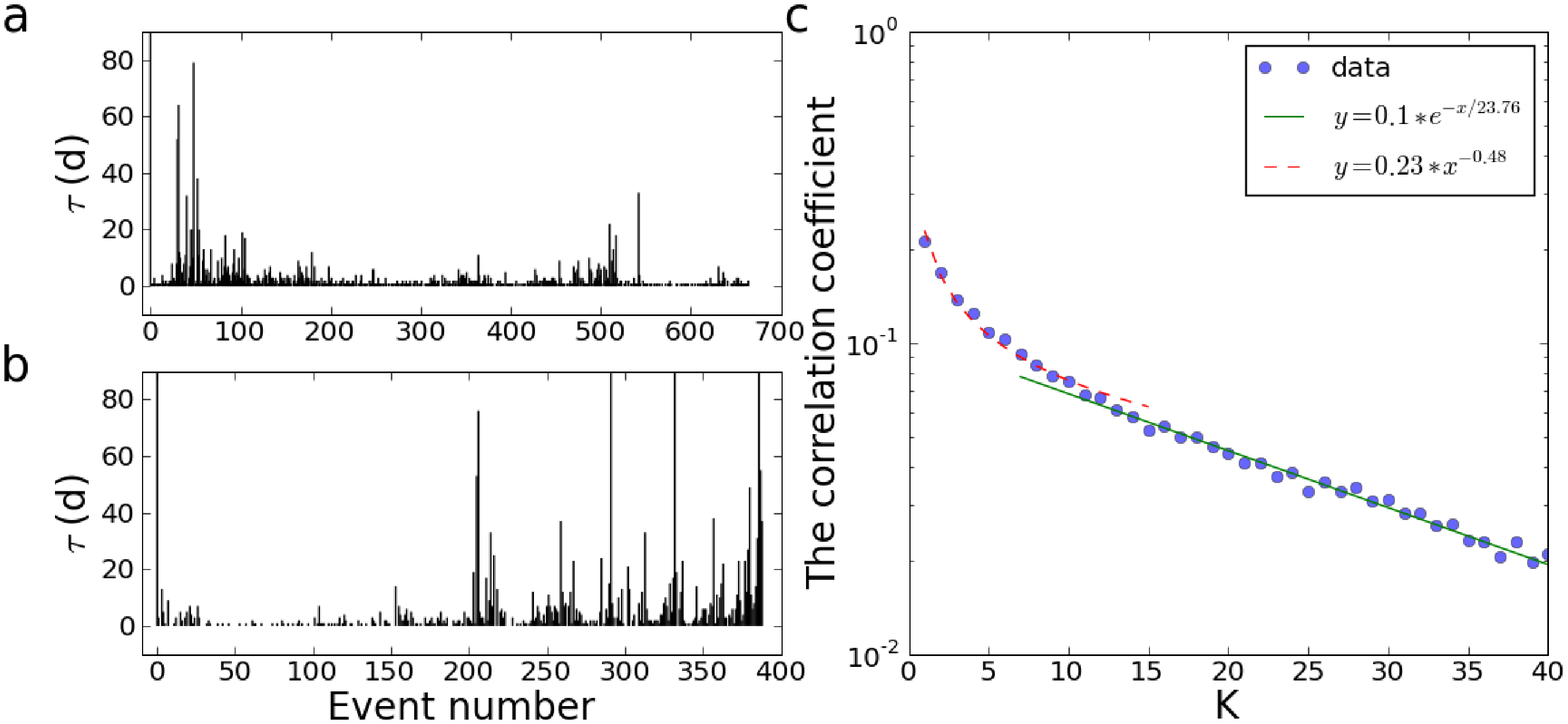}}
\caption{(Color online)(a), (b)The interevent time of consecutive
events of three users whose number of posts are 666,390.The user in
fig 3 is the same one in fig a; (c)The average of the memory
coefficient of all qualified users with different
$K$.}\label{afoto2}
\end{center}
\end{figure*}

Following the way in \cite{activity1}, We sort users in an ascending
order of $Activity$ $A_i$: $A_i=n_i/d_i$, where $n_i$ is the total
number of posts of user $i$ and $d_i$ is the time between the first
and the last post. For simplicity, we only divide these users into
two groups: one is the top 14000 users in the list above and the
average of $Activity$ $\langle A \rangle=0.04$ per day, the other
one is the remainder containing 6379 users and $\langle A
\rangle=0.64$ per day. As we can see in fig 3, the decay exponent of
interevent times distribution in group level increases from 1.86 to
2.47 as the $Activity$ from 0.04 to 0.64.

For individual behavior, we only consider users whose number of
posts is more than 200 to avoid characterizing users who post too
little. There are 2211 qualified users. Firstly, we choose one user
for example. As shown in fig 4, both the distribution and the
cumulative one of interevent times decay asymptotically as a power
law. The exponents of the cumulative distribution of this user is
1.40. Correspondingly, the one of distribution of this user are
2.40. The average of the exponents of all qualified users is 2.23.
The interevent times for consecutive events of this user is shown on
fig 5a. As comparison, the fig 5b is another user's. It can give us
the light of the human behavior from a visual understanding. One of
important features is the clustering of extreme long interevent
times which also is called mountain-valley-structure and can be
found in many complex systems\cite{mountain1,mountain2}. The
long-term change also can be found in these users: for fig 5b, the
posting frequency is obviously lower in the span of event number
200-400.

The features above inspire us to investigate the memory coefficient
of this succession, although the lack of it in human activities was
already reported by \cite{correlation1}. The definition of it is as
follow\cite{correlation1}:

\begin{equation}
M_k = \frac{1} {{{n_\tau } - 1}}\sum\limits_{i = 1}^{{n_\tau } - 1}
{\frac{{({\tau _i} - {m_1})({\tau _{i + k}} - {m_2})}} {{{\sigma
_1}{\sigma _2}}}},
\end{equation}where $\tau_i$ is the interevent time values and $n_\tau$ is the
number of interevent time and $m_1(m_2)$ and $\sigma _1(\sigma _1)$
are sample mean and sample standard deviation of $\tau_i$'s
($\tau_{i+k}$'s). The two interevent times $\tau_i$' and
$\tau_{i+k}$ is separated by k events.

Here, we calculate $M_k$ of all these qualified users with k ranging
from 1 to 40. Because the number of posts of one single user is
still so small that the decay curve of $M_k$ of one user presents
too big fluctuation, we only study the average $M_k$ of all users.
In fig 5c, $M_1$ is 0.21 which shows there is strong memory between
the nearest interevent times. Interestingly, there are two regimes
in this decay curve: when $k<10$, it decays asymptotically as a
power law: $y=0.23*x^{-0.48}$; when $k>10$, it decreased
exponentially: $y=0.1*e^{-x/23.76}$. This feature shows that the
short-term  and long-term memory in this behavior are likely due to
different mechanisms. For the part of $k>10$, the decay curve
reminds us of the Ebbinghaus forgetting curve which also has
exponential nature. It is possible that the exponential decay of
memory of our behavior lead to the same decay of the long-term
memory. For the part of $k<10$, it is much stronger and obviously
has something important to do with the mountain-valley-structure
above and even the heavy-tails in the distribution.

To sum up, there are three important features in this behavior: the
heavy-tails distribution of interevent time with an exponent $\beta
\approx 2$; the dependence between the exponent and $Activity$; the
strong memory. It is obvious that the stochastic models, such as the
priority-queue model\cite{dm7,dm8} and the cascading nonhomogeneous
poisson process\cite{dm1,dm2}, can't be the mechanism of this
correlated human dynamic. As to pervious memory-based models, the
adaptive interest model\cite{m2} only can give a distribution with
the exponent $\beta=1$; the exponent from Vazquez's model can more
than 2 \cite{m4}. However, two other features is absent in the
discussion of the both two models and the details, "overmany long
intervals", also can't get an explanation from them. Below we will
try to suppose a very simple model to explain all characteristics
above.

\section{Model and Simulation Results}
Before building our model, let's try to gain some intuition from our
daily life firstly. After along day's work, we have some free time
and need to do something that can relax ourselves. There are always
many choices for us: we can do exercises outside, or see a movie, or
listen to music, or write some words in your website like blogging,
and so on. In most cases, we would make a choice among them basing
on our personal preferences which are very diverse and different
from each other. However, in one way people are alike: the more
someone get interested in it, the more frequently he would do. On
the other hand, in a few cases, people would like to have a change
and try to do something new or something that wasn't done for a long
time.

So we assume that there are $N$ choices which can be regarded as
different forms of entertainment. At each time step, the agent
select one of them to do according to two "choosing rules" as
follows:

(1)Suppose one of $N$ choices was selected $T$ times in past $M$
steps, then the probability of choosing this one at current time
step is $T/M$.

(2)Picking up randomly from these $N$ choices

\begin{figure*}[ht]
\begin{center}
\centerline{\includegraphics[width=.6\textwidth]{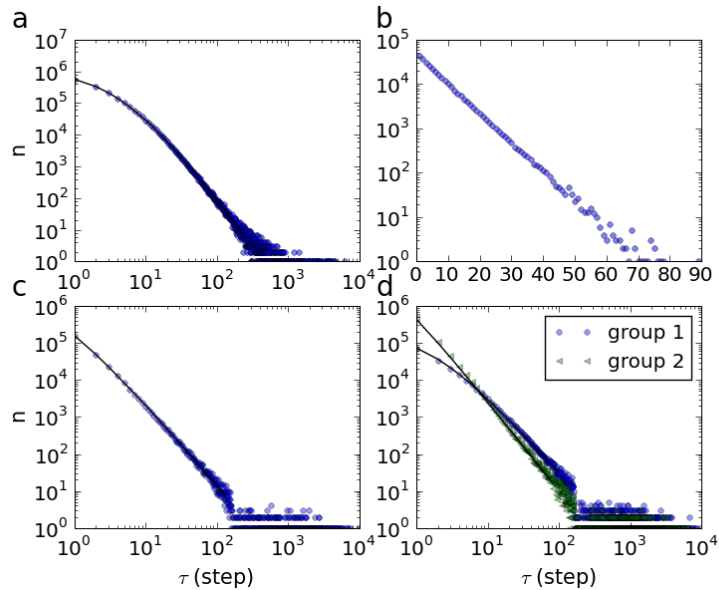}}
\caption{(Color online) The distribution of the interevent times
given by our simulation. In our simulation , $N=6$, $R=0.005$,(the
same in fig 6). we only observed the selection of one of these $N$
choices and the other's was ignored. At first, we make M choices
randomly as our initial conditions. Our program ran 2,000,000 steps
to get a stable exponent of the distribution. The result of
simulations when $M=1000$, $M=20000$, and $M=160$ is shown on (a),
(b), (c). For (a), the exponent $\beta = 2.92$ and $a=4.0$; the
distribution in (b) is obviously close to exponential one; For (c),
the exponent $\beta=2.19$ and $a=0.4$. The distribution of the
interevent time of the two groups with the same parameters of (c) is
shown on (d). For group 1, $\langle A_1 \rangle=0.055$, $\beta=2.10$
and $a=1.5$; for group 2, $\langle A_2 \rangle=0.295$, $\beta=2.41$
and $a=0.2$.}\label{afoto2}
\end{center}
\end{figure*}

The probability of executing the second rule is $R$ and the one of
executing the first rule is $1-R$. Here, the interevent times $\tau$
is the number of steps between choosing the same one consecutively.
From the description above, the probability of choosing $i$ in
current steps $P_i$ is:

\begin{equation}
{P_i} = (1 - R)T_i/M + R/N.
\end{equation}where $T_i$ is the times of selecting i in past $M$ steps.

In all simulations of this paper, $N$ is fixed to 6 which means that
we ignore new hobbies. It is obvious that the distribution of
interevent times approaches an exponential one when $R$ become
rather lager. So $R$ must be small but not be too small. The result
of following the preference rule totally is that the agent will only
select one of them repeatedly and ignore the others. In our
simulations, we suppose $R=0.005$. In $M\rightarrow\infty$ limit,
the distribution of interevent times become also exponential(seeing
fig 5b). It show this preference must be temporal, otherwise there
would be no heavy-tails! And $M$, as the most important parameters
in our model, show how temporal this preference is. The distribution
of the interevent time with $M=1000,20000,160$ was shown in fig 6.
As we can see, the exponent of this distribution when $M=160$ is
2.19 which is in agreement with the previous empirical values,
therefore $M$ of simulations blew defaults to 160.

It is noteworthy that the distribution in our case also has the
similar "overmany long intervals" in fig 2b and fig 4. Actually the
interevent time $\tau$ which is lager than $M$ is randomly generated
by the second 'choosing rule'. So for the part of $\tau>M$ it
deviate the power-law (for fig 6a and 6c, see the part of
$\tau>160$). But according to our data analysis it seems that this
extra long return times only can be found in loyal users who already
have posted a certain number of articles and most of users would
just abandon their blogs after suspending update for a long
time(seeing fig 2). In our simulation, basing on the second
'choosing rule', we actually assumed that the agent is a "super
loyal user" who would go on no matter how long he pause. That would
be why the distribution given by our simulation have so many long
intervals.

In order to get enough samples, we ran our program 100 times and
200,000 steps at each time. The selection made in last 60,000 steps
was recorded as one individual did. So we got 100 series which
belong to 100 different "users". Then we calculated the $A_i$ of
each "user" and sorted them by it. We choose the top 60 "users" as
one group whose $\langle A \rangle$ is 0.055 per step; the other
group contain the remaining 40 "users" and the $\langle A \rangle$
of it is 0.295 per step. From fig 6d, we can see similar dependence
between the exponent and $Activity$: the exponent $\beta$ increase
from 2.10 to 2.41 as the $Activity$ from 0.055 to 0.295.

In fig 7, the corresponding memory coefficient of our simulation
decays as a power law which can fit well the short-term memory
obtained from the data. For the long-term one or the part of $K>10$,
the coefficient of real data goes down exponentially with $K$ and is
about 0.01 smaller than the one of our model. The cause of this
discrepancy for long-term part may be that we do not take into
consideration the memory fading. It embodies in two aspects: for the
first choosing rule, the influence of past selection within $M$ time
steps is the same but in real behavior it should be decay with time;
on the other hand, we actually assumed that people would never
change his hobbies and be just trap in the $N$ choices and in real
life there is always a possibility of finding new hobby and abandon
old one. However, in our opinion, the discrepancy is too small to
affect the production of heavy-tails of this behaviors. The
short-term memory, which is much stronger and play a key role in the
origin of heavy-tails, is reproduced well by our model.

\begin{figure}[htb]
\includegraphics[width=0.4\textwidth]{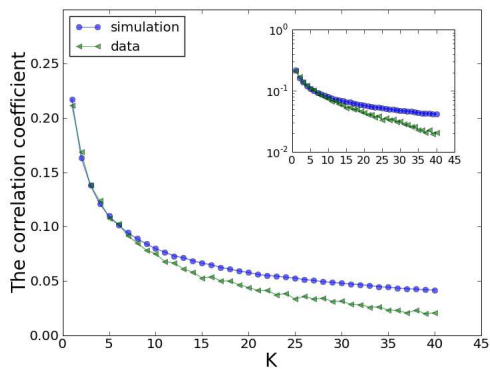}
\caption{\label{Fig:RI:PDF}(Color online) The comparsion between the
memory coefficient of the empirical result and simulation. Here,
with the same parameters as fig 5c, we perform 1000 simulations
which run 2000000 steps each time and get average $M_k$.}
\end{figure}

\section{Discussion}
There is high complexity in human behaviors. Different activities
can be conducted in different behavior pattern and the same activity
can be affect by multiple factors. To fully understand the pattern
of human behaviors we must investigate a wider variety of activities
and further details, not just the exponent of interevent time
distribution. In this paper, we investigated not only the
heavy-tails in the activity pattern of blog-posting but also the
memory and the role of $Activity$. In our opinion, it show there is
another kind of activities which also have the similar heavy-tails
nature but different origin. Although the influence of the seasonal
cycles also would be found in this behavior, the short-term memory
can't be explained by the model like the nonhomogeneous poisson
process\cite{dm1,dm2}.

One interesting result is that the decay curve of $M_k$ have two
regimes: for the short-term part($K<10$), it decays as a power law;
for the long-term part($K>10$), it decreases exponentially. Based on
the personal preferences rule: "the more we do it recently, the more
likely we will do it next", numerical simulations give the strong
short-term memory but the mechanism for long-term change is absent.
That would be why $M_k$ in the real date is a little smaller than
the one of our simulation for the $K>10$ part. Recent study also
pays attention to this kind of long-term change in human activity
\cite{dm2}. At the level of individual, this change can be the
result of personal interest and need shifts which seems very
unpredictable. However, at the level of population, the exponential
decay of memory hint that it have something to do with the memory
fading.

Another feature reproduced by our model is the dependence between
the $Activity$ and the exponent of the interevent time distribution.
According to the first rule of our model, the selective probability
is actually positive-linear dependent on the temporal $Activity$
$T_i/M$. That would be the cause of this dependence. Our model also
imply that a symbiotic relationship exists between the strong
short-term memory and this dependence. Further investigations are
needs in this direction.

Due to the complexity of our issue it is impractical to expect this
simple model to accurately match with the empirical result. One
important kinds of extensions of this model would be to consider the
effect of the memory fading as we discussed above. It is still a
question how to do it. One easy way would be to assume that the
weight of influence of the past choices decay with time. However, is
there more nature ways? We mean to find the mechanism of memory and
figure out why the strength of memory decay exponentially. One
reason of forgetting the old hobby can be finding a new one. But the
detail process of it is still unknown. Interaction is another factor
needed to be into consideration. Human, as a social beings, live in
a network knitted by friends and relatives and cannot avoid the
effect from it. But interaction seems be secondary to the
heavy-tails as the interevent time distribution of some systems
without interaction also have the heavy-tails\cite{activity2} and
the model with interactions show it just increases the value of
exponent\cite{m3}. If the stimulation from friends can make people
more actively, our model actually include this effect naturally
since the exponent from our model also increases with the
$activity$.

In our opinion, this strong short-term memory correlation should be
common in the activities which is more a matter of personality than
task. We hope that more empirical studies would be made in the
future. Actually, people not only have preferences for different
activities, even for the same one with different types. Taking
watching movies for example, when we decide to watch a DVD, there
are usually many kinds of movies: romance, sci-fi, classics,
horror....in this case, the $N$ choices of our model can be regarded
as different types of movies. In the dynamic evolution of social
network, we also can treat one's friends as the choices of our model
when people try to choose one of their friends to contact. In our
opinion, the personal preferences can be found both in many
activities and at different levels.

The memory or correlation of human behavior have much to do with the
predictability. One's friends can understand and predict his
behaviors better because they know his past and one's past, present
and future are interconnected. A music that many people like is
probably liked by you since we are influence each other and
correlation could be found in the preferences of us. Revealing it in
human behaviors and uncovering the mechanism of correlated human
dynamics have great significance for many fields, such as the link
prediction and recommender systems. Taking the recommender systems
for example. All recommendation algorithms including the famous
"page-rank" algorithms used by Google is just based on the empirical
hypothesis. The practice shows it works, but why it works? And how
to find the best recommendation algorithm? Without fully
understanding human behavior, especially the relation between one's
past behavior and present , we can not answer these questions well.

\begin{acknowledgments}
This work is funded by the National Basic Research Program of China
(973 Program No.2006CB705500);The National Important Research
Project(Study on emergency management for non-conventional happened
thunderbolts, Grant No. 91024026);the National Natural Science
Foundation of China (Grant Nos. 10975126, 10635040); the Specialized
Research Fund for the Doctoral Program of Higher Education of
China(Grant No:20093402110032)
\end{acknowledgments}

\end{document}